\documentclass[aps,prl,twocolumn,groupedaddress,amsmath,amssymb]{revtex4-1}

\usepackage{color}

\usepackage{graphicx}
\usepackage{bm}
\usepackage{braket}

\begin{document}

\title{Quantum electrodynamics of a free particle near dispersive dielectric or conducting boundaries}

\author{Robert Bennett and Claudia Eberlein}

\affiliation{Department of Physics \& Astronomy, University of Sussex, Falmer, Brighton BN1 9QH, UK}

\date{\today}

\begin{abstract}
Quantum electrodynamics near a boundary is investigated by considering the inertial mass shift of an electron near a dielectric or conducting surface.  We show that in all tractable cases the shift can be written in terms of integrals over the TE and TM reflection coefficients associated with the surface, in analogy to the Lifshitz formula for the Casimir effect. We discuss the applications and potential limitations of this formula, and provide exact results for several models of the surface.  \end{abstract}


\maketitle

Quantum electrodynamics is the spectacularly successful theory of the interaction between charges and electromagnetic fields. The anomalous magnetic moment of the electron is perhaps its famous result, finding agreement between theory and experiment to an accuracy of one part in $10^{13}$ \cite{Gabrielse,Hanneke}. However, no conceivable experiment can measure a quantity in isolation - there will always be apparatus and environment-dependent effects \cite{Fischbach, Svozil, Bordag, BoulwareBrown, KreuzerSvozil, Kreuzer, BartonFawcett}. One type of correction arises from modification of the quantized electromagnetic field due to material boundaries in the vicinity of the system under consideration. Here, we consider the effect that the modified quantized field has on the self-energy of an electron. While the corresponding calculations of other quantities such as the magnetic moment shift \cite{MagMomPaper} have more obvious experimental relevance, the mass shift calculation turns out to be technically simple, and gives exact results even for dispersive surfaces. Thus, we present the self energy calculation as both an accessible example of our formulation of quantum field theory near boundaries, and as a clarification of a previous result \cite{EberleinRobaschik, PRDEberleinRobaschik}.

To find the mass shift using the standard formalism of quantum field theory, one calculates the self-energy diagram to the respective order of interest. To 1-loop order $e^2\equiv\alpha$ and in free space, this is a straightforward calculation (cf. \cite{Peskin}). But, when boundaries are present even 1-loop calculations of quantum electrodynamics get very cumbersome \cite{PRDEberleinRobaschik}. Since we seek only the change in the self energy that is attributable to the surface, we take a different, more appropriate approach.  While the photon propagator receives boundary dependent corrections, the electron propagator does not (provided the electron is sufficiently far away from the boundary that there is no wave-function overlap with the structure of the surface, and the interaction between the particle and the surface is purely electromagnetic). This means a Feynman diagrammatic approach to the boundary dependent shift is not in fact necessary, and it suffices to study a first-quantized electron interacting with a second-quantized photon field.

We consider a material filling the space $z>0$, described by some dielectric function $\epsilon(\omega)$. The electron sits in vacuum, a distance $z=-|z|$ away from the material. Via reflection and refraction, the presence of the surface affects the electromagnetic field, which interacts with the electron via  \footnote{We work in natural units $c = 1 = \hbar$, $\epsilon_0 = 1$.}:

\begin{equation}
H_{\text{int}} = -\frac{e}{m}\mathbf{p} \cdot \mathbf{A} + V_{\text{image}}
\end{equation}

where $V_{\text{image}}$ is the electrostatic image potential for a non-dispersive medium. As shown in \cite{BartonPlasma1977}, this interaction Hamiltonian remains valid for dispersive media modelled as a plasma, via the use of a particular unitary transformation. Second-order perturbation theory then gives the self energy as

\begin{equation}
\Delta E = \frac{e^2}{m^2} \sum_{k,\lambda} \sum_{p_f} \frac{|\bra{p_f;1_{k \lambda}}\mathbf{p}\cdot  \mathbf{A} \ket{p;0}|^2}{\frac{\mathbf{p}^2}{2m}-\left[\frac{\mathbf{p}_f^2}{2m} +  \omega\right]}
\end{equation}

for a photon of momentum $k$ and polarization $\lambda$. We make the no-recoil approximation and take the electron's final momentum $p_f$ to be equal to its initial momentum $p$, which is a reasonable assumption to make as we are dealing with a low-energy effect. Writing the quantized field $\mathbf{A}$ in terms of mode functions $\mathbf{f}_{\mathbf{k}\lambda}$ and standard photon creation and annihilation operators via

\begin{equation}
\mathbf{A} = \sum_{\text{all modes}} (\mathbf{f}_{\mathbf{k}\lambda} a_{\mathbf{k} \lambda}+\mathbf{f}_{\mathbf{k}\lambda}^* a_{\mathbf{k} \lambda}^\dagger) ,
\end{equation}

we can express the self-energy in the form

\begin{equation}
\Delta E = -\frac{e^2}{m^2} \int d^3 k \sum_{\lambda} \frac{1}{\omega_\lambda}\left(|f_\parallel^\lambda|^2\langle p_\parallel^2 \rangle + |f_z^\lambda|^2\langle p_z^2 \rangle\right).
\end{equation}

We initially consider two choices of dielectric functions. These are a non-dispersive dielectric with $\epsilon(\omega) = n^2$, and an undamped plasma with dielectric function

\begin{equation}
\epsilon(\omega) = 1-\frac{\omega_p^2}{\omega^2},\label{PlasmaDielectricFunction}
\end{equation}

where $\omega_p$ is the plasma frequency. The modes $\mathbf{f}_{\mathbf{k}\lambda} $ can be written in terms of plane waves with reflection and transmission coefficients $R_\lambda$ and $T_\lambda$ relevant to incidence from either side of the interface. These coefficients are given by the standard Fresnel expressions. Writing wave vectors as $\mathbf{k} = (k_\parallel,k_z)$ on the vacuum side and $(k_\parallel,k_z^d)$ on the medium side, the Fresnel coefficients that turn out to be of importance are the reflection coefficients for left-incident radiation

\begin{equation}
R^L_{TE} = \frac{k_z - k_z^d}{k_z + k_z^d} \qquad R^L_{TM} = \frac{\epsilon(\omega) k_z - k_z^d}{\epsilon(\omega) k_z + k_z^d}  \label{RCoeffs}\\
\end{equation}

where  $k_z^d = \sqrt{\epsilon(\omega)(k_z^2+k_\parallel^2)-k_\parallel^2}$, and we choose $\text{sgn}(k_z) = \text{sgn}(k_z^d)$. The mode functions in the non-dispersive case are:

\begin{widetext}
\begin{align}
f_{k \lambda,\text{nondisp}}^{L} &= \frac{1}{(2\pi)^{3/2}}\frac{1}{\sqrt{2\omega}} \left\{\theta(-z)[e^{i\mathbf{k}\cdot \mathbf{r}} \hat{e}_\lambda(\mathbf{k})+R^L_\lambda e^{i\bar{\mathbf{k}}\cdot \mathbf{r}} \hat{e}_\lambda(\bar{\mathbf{{k}}})] + \theta(z)T^L_\lambda e^{i\mathbf{k}^d\cdot \mathbf{r}} \hat{e}_\lambda(\mathbf{k}^d) \right\} \notag \\
f_{k\lambda,\text{nondisp}}^{R} &=\frac{1}{(2\pi)^{3/2}} \frac{1}{\sqrt{2\omega}} \frac{1}{n}\left\{\theta(z)[e^{i\mathbf{k}^d\cdot \mathbf{r}} \hat{e}_\lambda(\mathbf{k}^d)+R^R_\lambda e^{i\bar{\mathbf{k}}^{d}\cdot \mathbf{r}} \hat{e}_\lambda(\bar{\mathbf{k}}^{d})] + \theta(-z)T^R_\lambda e^{i\mathbf{k}\cdot \mathbf{r}} \hat{e}_\lambda(\mathbf{k}) \right\}   \label{NonDispModes}
\end{align}

where the modes have been split according to left and right incidence. Barred $k$ vectors correspond to modes that have undergone a reflection (reversing the sign of their $z$ component). The modes for the plasma surface are obtained from these by the replacements

\begin{equation*}
f_{k \lambda,\text{plasma}}^{L} = f_{k \lambda,\text{nondisp}}^{L}(n^2 \to \epsilon(\omega)) \qquad, \qquad  f_{k \lambda,\text{plasma}}^{R} = n f_{k \lambda,\text{nondisp}}^{R}(n^2 \to \epsilon(\omega)) 
\end{equation*}

and the addition of a new surface-plasmon mode function

\begin{align}
f_{k,sp} &= \frac{1}{2\pi} \frac{1}{\sqrt{p(k)}}\left(\Theta(-z)\left(\hat{k}_\parallel -i(k_\parallel/\kappa) \hat{z}\right) e^{i \mathbf{k}_\parallel \cdot \mathbf{r}_\parallel+ \kappa z}+\Theta(z)\left(\hat{k}_\parallel +i({k_\parallel}/{\kappa^d}) \hat{z}\right) e^{i \mathbf{k}_\parallel \cdot \mathbf{r}_\parallel- \kappa^d z}\right) \label{Modes}
\end{align}

where we have defined the norming function $p(k)=(\epsilon^4-1)/({\epsilon^2 \sqrt{-1-\epsilon}})$ and $\kappa^{(d)}>0$. We split the shift into contributions proportional to $\langle p_\parallel^2 \rangle$ and $\langle p_z^2 \rangle$, so that the energy shift is written as

\begin{equation}
\Delta E = \sum_\lambda \left(\Delta E_\parallel^{(\lambda)} + \Delta E_\perp^{(\lambda)}\right). 
\end{equation} 

Specializing initially to the non-dispersive case, the $k_z$ integrals arising in $\Delta E^{(\lambda)}_\perp$ (or $\Delta E^{(\lambda)}_{\parallel}$) can be written in the form

\begin{align*}
& \left\{ \int_0^\infty dk_z \,  [1+|R^L_\lambda|^2 \pm  R^L_\lambda(e^{2ik_zz}+e^{-2ik_zz})]  +\frac{1}{n^2} \int_{-\infty}^{-\Gamma} dk_z^d  |T^R_\lambda|^2 \pm \frac{1}{n^2} \int_{-\Gamma}^0 dk_z^d  |T^R_\lambda|^2 e^{2|k_z| z} \right\}\frac{g_\lambda(k_\parallel, k_z)}{(k_z^2+k_\parallel^2)^2} \\
  =&  \left\{\int_0^\infty dk_z \,  [1+|R^L_\lambda|^2] +\frac{1}{n^2} \int_{-\infty}^{-\Gamma} dk_z^d  |T^R_\lambda|^2  \pm \left[\int_0^\infty dk_z \,   R^L_\lambda(e^{2ik_zz}+e^{-2ik_zz})] + \frac{1}{n^2} \int_{-\Gamma}^0 dk_z^d |T^R_\lambda|^2 e^{2|k_z| z} \right]\right\}\frac{g_\lambda(k_\parallel, k_z)}{(k_z^2+k_\parallel^2)^2}
\end{align*}
\end{widetext}

where $\Gamma = \sqrt{n^2-1}k_\parallel$ and  $g_\lambda(k_\parallel, k_z)$ is a function analytic in $k_z$, and which is specific to each polarization. The three integrals in the first line represent the left incident modes, the right-incident travelling modes and the right-incident evanescent modes, respectively. The first two integrals in the second line are the same as they would be in free space, whence we subtract them, since we are interested only in the boundary-dependent part of the shift. The second two integrals (those in the square brackets) can be combined into a single contour integral by observing that for $k_z$ in the upper half-plane

\begin{equation}
R^L_{\lambda} |_{ k_z^d = -K} - R^L_{\lambda}|_{k_z^d = K} = \frac{k_z}{k_z^d} T^R_\lambda T^{R*}_\lambda |_{ k_z^d = -K},
\end{equation}

where $K>0$, meaning the two integrals can be combined into 

\begin{equation}
\pm  \int_C dk_z \,\frac{g_\lambda(k_\parallel, k_z)}{(k_z^2+k_\parallel^2)^2}  R^L_\lambda e^{-2ik_z z} .
\end{equation}

The contour $C$ runs over the interval $[-\infty,0]$, up the imaginary axis to the point $ik_\parallel \sqrt{n^2-1}/n$, back down to the real axis and then along $[0,-\infty]$. Taking the same approach for the plasma surface, the contour in this case is simply $[-\infty, \infty]$, passing over a branch cut from $-\omega_p$ to $\omega_p$. Deforming these contours into the upper half-plane results in an additional contribution from a pole at zero frequency, and for the plasma case, a contribution from a pole in the TM reflection coefficient. We find that this latter contribution exactly cancels the surface plasmon part of the mode functions (as seen in \cite{BabikerPlasma,MagMomPaper}). Thus, we can write the result for either model in terms of a single formula over reflection coefficients, with the choice of surface entering only via choice of $\epsilon(\omega)$ in equation (\ref{RCoeffs}). These integrals are:

\begin{widetext}

\begin{align}
\Delta E_\parallel &=\frac{e^2}{32\pi m^2}\langle p_\parallel^2 \rangle\int_0^\infty d k_\parallel k_\parallel   e^{2k_\parallel z} \left[-\frac{2R^L_{TE}}{k_\parallel} +i \frac{d R_{TM}^L}{dk_z} +R_{TM}^L  \left(2z+  \frac{1}{k_\parallel}\right)\right] \Bigg|_{k_z \to ik_\parallel} \label{DeltaEPara} \\
\Delta E_{\perp}&=\frac{e^2 }{16\pi m^2 }\langle p_z^2 \rangle\int_0^\infty d k_\parallel k_\parallel e^{2 k_\parallel z}  \left[i \frac{d R_{TM}^L}{dk_z} - R_{TM}^L\left(\frac{1}{  k_\parallel}-2 z \right) \right]\Bigg|_{k_z \to ik_\parallel}  \label{DeltaEPerp}.
\end{align}
\end{widetext}

These formulae arise from residues of a double pole, but the dependence on the reflection coefficients is written as explicitly as possible for later convenience. To evaluate the results for the two models, one simply inserts the relevant dielectric function into the above formulae. However, we first consider the case of perfect reflectivity -- long held to be a good-enough approximation to real surfaces. To effect this we take the $n^2\to\infty$ limit of the reflection coefficients, resulting in $R^L_{TE}=-1$, $R^L_{TM}=1$. The integrals are trivial, and the result is

\begin{equation}
\Delta E_{\text{PM}} = -\frac{e^2}{32m^2 \pi  z } \langle p_\parallel^2 \rangle  +\frac{e^2}{16\pi m^2 z}     \langle p_z^2 \rangle 
\end{equation}

in agreement with \cite{EberleinRobaschik}. Now using $\epsilon(\omega) = n^2$ in equation (\ref{RCoeffs}), we find

\begin{align}
\Delta E_{\text{NonDisp}} =& \frac{e^2}{32m^2 \pi  z }\frac{n^2 \left(n^2-1\right)}{\left(1+n^2\right)^2 } \langle p_\parallel^2 \rangle \notag \\
& +\frac{e^2}{16\pi m^2 z}  \frac{2n^4 -n^2 -1 }{(n^2+1)^2}    \langle p_z^2 \rangle  .
\end{align}

The $n\to\infty$ limit of this result clearly does not agree with the perfect reflector, as discussed in \cite{EberleinRobaschik}. This discrepancy is one of a family of such issues, we discuss the physical meaning and origin of these after presenting further results.

Proceeding, we insert the plasma dielectric function (\ref{PlasmaDielectricFunction}) into (\ref{RCoeffs}) and carry out the integrals (\ref{DeltaEPara}) and (\ref{DeltaEPerp}). Noting that the $k_z \to ik_\parallel$ limits of the plasma reflection coefficients coincide with the reflection coefficients for the perfect mirror, we find that we can write the plasma shifts as corrections to the perfect mirror case via:

\begin{align}
\Delta E_{\perp\text{,Plasma}} &= \Delta E_{\perp,PM} + \Delta_\perp \label{PlasmaShiftPerp}\\
\Delta E_{\parallel\text{,Plasma}} &= \Delta E_{\parallel,PM} + \Delta_\parallel \label{PlasmaShiftPara}
\end{align}
with  
\begin{align*}
\Delta_\perp &= -\frac{e^2}{4 \pi m^2 \omega_p^2}\langle p_z^2 \rangle\int_0^\infty dk_\parallel k_\parallel \sqrt{k_\parallel^2+\omega_p^2}e^{2 k_\parallel z} \\
\Delta_\parallel &=  -\frac{e^2}{4 \pi m^2 \omega_p^2}\langle p_\parallel^2 \rangle\int_0^\infty dk_\parallel k_\parallel\left( \sqrt{k_\parallel^2+\omega_p^2}-\frac{k_\parallel}{2}\right)e^{2 k_\parallel z}.
\end{align*}

The integrals $\Delta_{\perp,\parallel}$ can be evaluated analytically, but the resulting expressions are lengthy combinations of special functions so we will not quote them here. The limiting case $\omega_p \to\infty$ is
 \begin{equation*}
\lim_{\omega_p\to \infty} \Delta_{\perp,\parallel} =0,
\end{equation*}

showing that the plasma and perfect mirror models are equivalent at $\omega_p\to\infty$. 

The surface models discussed so far have all had a desirable common feature, namely that their corresponding wave equations are all Hermitian eigenvalue problems (cf. \cite{GlauberLewenstein}). A first-principles derivation of the mass shift using a dielectric function which does not have this property is not possible since one cannot derive the modes. However, one can use the Lifshitz theory and write the electromagnetic field as a response to fluctuating noise currents inside the material (see, for example, \cite{JenaReview}). The Green's function describing this response involves the reflection coefficients of the surface so that one necessarily ends up with the same formulae (\ref{DeltaEPara}) and (\ref{DeltaEPerp}), regardless of the specific choice of dielectric function. Proceeding along these lines, we use our formulae to investigate a model in which a restoring force is introduced \cite{BartonPlasma}, summarized by the introduction of a new characteristic frequency $\omega_T$ into the dielectric function:

\begin{equation}
\epsilon(\omega) = 1- \frac{\omega_p^2}{\omega^2-\omega_T^2} \label{DispDielFunction}.
\end{equation} 

Surprisingly, the integrals are here much simpler than for the plasma case, the result is:

\begin{widetext}
\begin{equation*}
\Delta E_{\text{Disp}} = \frac{e^2\omega_p^2}{16\pi m^2\left(\omega_p^2+2 \omega_T^2\right)^2}\left\{\left[\frac{1}{2z^3}+ \frac{1}{2z}\left( \omega_p^2+ \omega_T^2\right) \right] \langle p_\parallel^2 \rangle+\left[\frac{1}{z^3}+\frac{1}{z}(2 \omega_p^2+3 \omega_T^2) \right] \langle p_z^2 \rangle   \right\}\label{DeltaEDisp}.
\end{equation*}
\end{widetext}

We can compare this result to the non-dispersive case by using the static susceptibility $\chi(0)$

\begin{equation*}
\chi(0) = \epsilon(0)-1 =\begin{cases} n^2-1 &\mbox{non-dispersive } \\ 
\omega_p^2/\omega_T^2 & \mbox{dispersive}  \end{cases} 
\end{equation*}

This gives, for example, the perpendicular component as

\begin{equation*}
\Delta E_{\text{Disp},\perp} = \frac{e^2 }{16 \pi m^2 z  }\frac{\chi(0)}{(\omega_T z)^2}\frac{ 1+(\omega_T z)^2 (3+2 \chi(0) )}{ (2+\chi(0) )^2}\langle p_z^2 \rangle.
\end{equation*}

\begin{figure}[h]
\includegraphics{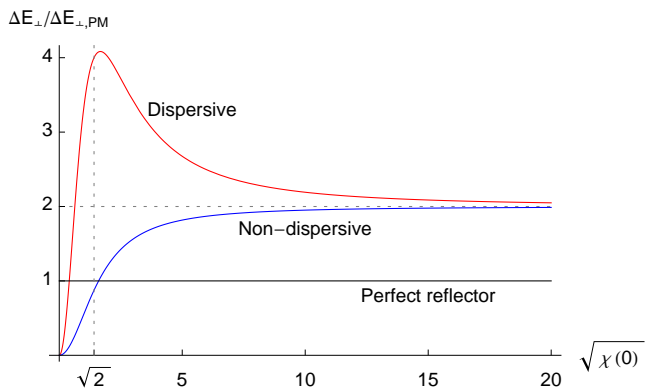}
\caption{[Colour online] Energy shift vs static susceptibility for various models in units of the perfect reflector shift. The position and height of the peak may be tuned by varying $\omega_T z$, and the factor-of-two disagreement between the perfect reflector and dielectric cases is shown.} \label{Graph}
\end{figure}

\begin{figure}[h]
\includegraphics{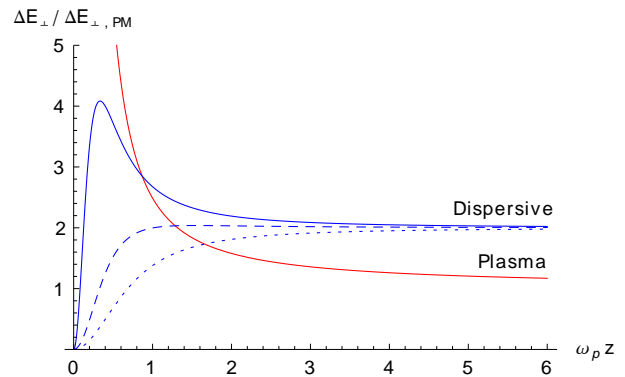}
\caption{[Colour online] Energy shift in units of the perfect reflector shift vs the dimensionless parameter $\omega_p z$ for the plasma and dispersive dielectric models for various $\omega_T z$. The values of $\omega_T z$ shown are $0.2$ (solid line), $0.4$ (dashed) and $0.6$ (dotted).} \label{AdditionalGraph}
\end{figure}

It is seen that, for large $\chi(0)$, the dispersive dielectric is in agreement with the non-dispersive dielectric with large $n$. Figure \ref{Graph} shows the energy shift as dependent on the static susceptibility for $\omega_T z = 0.2$. The plasma result cannot be shown in Fig.~(\ref{Graph}) since its static susceptibility is infinite; however it can be compared to the dispersive shift via a plot against the dimensionless parameter $\omega_p z$, as shown in Fig.~(\ref{AdditionalGraph}). In both plots we show energy shifts in units of the perfect reflector shift. 

It is not hard to show that the peak emerges only when $\omega_T z<1/	\sqrt{5}$. It moves towards $\chi(0)=2$ for decreasing $\omega_T z$, and its height scales as $(\omega_T z)^{-2}$. Thus, for small values of this $\omega_T z$ we see that the shift can be made considerably larger than in the previously considered perfect reflector model. 

Now we turn our attention to the origins of the discrepancies between the models. Mathematically, the disagreements arise because of non-commutation of limits in the reflection coefficients (or their derivatives), namely $k_z \to ik_\parallel$ (required for finding the residue at this point) and whatever limit one has to take to get from dielectric function to another. For example, the $\omega_T \to 0$ limit of the result for the dispersive dielectric should take us to the plasma result, but it does not. This is because the $\omega_T \to 0$ and $k_z\to ik_\parallel$ limits of the derivative of the TM reflection coefficient do not commute. A similar problem causes the perfect reflector and non-dispersive results to disagree in the limit $n\to\infty$. 

Physically, the differences between models that disagree with each other are down to a number of reasons. One of them is the exclusion of part of the photon phase-space, namely the evanescent modes. Previous workers have shown that exclusion of evanescent modes is not an adequate approximation to reality \cite{PRDEberleinRobaschik}, we confirm this conclusion in the context of a dispersive medium.  The other main reason has to do with the different response of conductors and dielectrics to electric fields at low frequencies: $\varepsilon(\omega)$ has a pole at $\omega=0$ for a conductor but not for a dielectric. The discrepancies between the results for the mass shift show that one has to decide whether the material at hand should be modelled as a metal (no restoring force for the charge carriers) or as a dielectric (with a restoring force parametrized by $\omega_T$), since these two classes of model for the surface are not obtainable as limiting cases of one another, reflecting the different nature of the electromagnetic response of conductors and dielectrics. 

To find conductors and dielectrics giving rise to different results on account of their different response to electromagnetic fields is of course not at all surprising. This is, however, in contrast to what one might have expected from the closely related Casimir-Polder energy shift in an atom close to a conducting or dielectric boundary: in both retarded and non-retarded regimes the Casimir-Polder shift of an atom in front of a dielectric \cite{Wu} reproduces the original result for an atom close to a perfect reflector \cite{CP} in the limit of infinite dielectric constant, and so does the level shift for an atom near a plasma surface \cite{BabikerBarton} in the limit of infinite plasma frequency, $\omega_p\longrightarrow\infty$. The crucial difference between an atom and a free particle in this context is that the excitation spectrum of a bound electron has a gap at low frequencies corresponding to the nearest energy level whereas a free particle admit excitations of arbitrarily low frequency. As a consequence, the low-frequency behaviour of the electromagnetic response of the material, in particular the pole at $\omega=0$ in the dielectric function of a conductor, play a decisive role for the mass shift of a free particle, but not for the Casimir-Polder shift of an atom. 

The decisive importance of the pole at $\omega=0$ in the dielectric function of a conductor is made obvious by the fact that the energy shifts (\ref{PlasmaShiftPerp}) and (\ref{PlasmaShiftPara}) do not vanish in the limit $\omega_p\to 0$, despite $\epsilon(\omega)$ reducing to the vacuum value of $1$ in that case. The limit $\omega_p\to 0$ is non-analytic because the choice of a dielectric function of the form (\ref{PlasmaDielectricFunction}) necessarily describes freely moving charge carriers at $\omega = 0$, which is obviously not true for vacuum with $\epsilon \equiv 1$. Mathematically speaking, eq.~(\ref{PlasmaDielectricFunction}) is ill-defined if both $\omega \to 0$ and $\omega_p \to 0$; in line with the physical interpretation, the fact that $\epsilon(\omega)$ has a pole at $\omega = 0$ is more important than the strength of this pole. 

A natural next step in the investigation of the effect for realistic materials would be to include a damping parameter, meaning that the dielectric function in terms of $k_z$ is:
\begin{equation}
\epsilon(k_\parallel, k_z)  = 1-\frac{\omega_p^2}{\sqrt{k_z^2+k_\parallel^2}\left(\sqrt{k_z^2+k_\parallel^2}+i \gamma\right)}.\label{DampedDielectricFunction}
\end{equation}
This introduces the additional complication that the reflection coefficient has branch points at $k_z =\pm ik_\parallel$, causing the formulae (\ref{DeltaEPara}) and (\ref{DeltaEPerp}) to become ambiguous. Our method is reliant on the fact that whatever happens below $ik_\parallel$ in the upper half of the $k_z$ plane does not preclude a contour deformation that allows us to simply pick up the residue at $k_z=ik_\parallel$. This inability to deform a contour once damping is introduced has been investigated in detail in \cite{BordagDrude}, where it is concluded that standard derivations of the Lifshitz formula with damping must contain an unspecified inconsistency (a statement reflected by experimental results \cite{DeccaCasimirPressure}). In our approach, the failure of the damped reflection coefficient to be single-valued in the neighborhood of $k_z=ik_\parallel$ could be a manifestation of these problems with damping detailed by others. 

Our results are intimately related to the shift in the cyclotron frequency of an electron near a surface. If the external magnetic field is directed perpendicular to the surface, the calculations coincide, so that a measurement of the cyclotron frequency is in effect a measurement of the mass shift. Precision $g-2$ experiments often rely on accurate measurement of the cyclotron frequency \cite{Gabrielse,Hanneke}, but a much more elaborate calculation would have to be undertaken in order to precisely enumerate the effects relevant to a specific apparatus. Finally we note that for magnetic fields directed parallel to the surface, the additional electrostatic interaction skews the orbit, and much more so than the mass anisotropy \cite{BartonFawcett}. Thus, a measurement of the cyclotron frequency in a parallel field does not deliver the mass shift. 

In summary, we have calculated the self-energy of an electron near dispersive surfaces of various kinds, which had previously been considered only in idealized models. We have shown that is it crucial that one decides at the start of a calculation whether the material should be modelled as a metal or a dielectric, since the results for the two classes of material are not obtainable as limiting cases of one another.

It is a pleasure to thank Robert Zietal for discussions. Financial support from the UK Engineering \& Physical Sciences Research Council is gratefully acknowledged.


\begin{thebibliography}{24}%
\makeatletter
\providecommand \@ifxundefined [1]{%
 \@ifx{#1\undefined}
}%
\providecommand \@ifnum [1]{%
 \ifnum #1\expandafter \@firstoftwo
 \else \expandafter \@secondoftwo
 \fi
}%
\providecommand \@ifx [1]{%
 \ifx #1\expandafter \@firstoftwo
 \else \expandafter \@secondoftwo
 \fi
}%
\providecommand \natexlab [1]{#1}%
\providecommand \enquote  [1]{``#1''}%
\providecommand \bibnamefont  [1]{#1}%
\providecommand \bibfnamefont [1]{#1}%
\providecommand \citenamefont [1]{#1}%
\providecommand \href@noop [0]{\@secondoftwo}%
\providecommand \href [0]{\begingroup \@sanitize@url \@href}%
\providecommand \@href[1]{\@@startlink{#1}\@@href}%
\providecommand \@@href[1]{\endgroup#1\@@endlink}%
\providecommand \@sanitize@url [0]{\catcode `\\12\catcode `\$12\catcode
  `\&12\catcode `\#12\catcode `\^12\catcode `\_12\catcode `\%12\relax}%
\providecommand \@@startlink[1]{}%
\providecommand \@@endlink[0]{}%
\providecommand \url  [0]{\begingroup\@sanitize@url \@url }%
\providecommand \@url [1]{\endgroup\@href {#1}{\urlprefix }}%
\providecommand \urlprefix  [0]{URL }%
\providecommand \Eprint [0]{\href }%
\providecommand \doibase [0]{http://dx.doi.org/}%
\providecommand \selectlanguage [0]{\@gobble}%
\providecommand \bibinfo  [0]{\@secondoftwo}%
\providecommand \bibfield  [0]{\@secondoftwo}%
\providecommand \translation [1]{[#1]}%
\providecommand \BibitemOpen [0]{}%
\providecommand \bibitemStop [0]{}%
\providecommand \bibitemNoStop [0]{.\EOS\space}%
\providecommand \EOS [0]{\spacefactor3000\relax}%
\providecommand \BibitemShut  [1]{\csname bibitem#1\endcsname}%
\let\auto@bib@innerbib\@empty
\bibitem [{\citenamefont {Odom}\ \emph {et~al.}(2006)\citenamefont {Odom},
  \citenamefont {Hanneke}, \citenamefont {D'Urso},\ and\ \citenamefont
  {Gabrielse}}]{Gabrielse}%
  \BibitemOpen
  \bibfield  {author} {\bibinfo {author} {\bibfnamefont {B.}~\bibnamefont
  {Odom}}, \bibinfo {author} {\bibfnamefont {D.}~\bibnamefont {Hanneke}},
  \bibinfo {author} {\bibfnamefont {B.}~\bibnamefont {D'Urso}}, \ and\ \bibinfo
  {author} {\bibfnamefont {G.}~\bibnamefont {Gabrielse}},\ }\href {\doibase
  10.1103/PhysRevLett.97.030801} {\bibfield  {journal} {\bibinfo  {journal}
  {Phys. Rev. Lett.}\ }\textbf {\bibinfo {volume} {97}},\ \bibinfo {pages}
  {030801} (\bibinfo {year} {2006})}\BibitemShut {NoStop}%
\bibitem [{\citenamefont {Hanneke}\ \emph {et~al.}(2008)\citenamefont
  {Hanneke}, \citenamefont {Fogwell},\ and\ \citenamefont
  {Gabrielse}}]{Hanneke}%
  \BibitemOpen
  \bibfield  {author} {\bibinfo {author} {\bibfnamefont {D.}~\bibnamefont
  {Hanneke}}, \bibinfo {author} {\bibfnamefont {S.}~\bibnamefont {Fogwell}}, \
  and\ \bibinfo {author} {\bibfnamefont {G.}~\bibnamefont {Gabrielse}},\ }\href
  {\doibase 10.1103/PhysRevLett.100.120801} {\bibfield  {journal} {\bibinfo
  {journal} {Phys. Rev. Lett.}\ }\textbf {\bibinfo {volume} {100}},\ \bibinfo
  {pages} {120801} (\bibinfo {year} {2008})}\BibitemShut {NoStop}%
\bibitem [{\citenamefont {Fischbach}\ and\ \citenamefont
  {Nakagawa}(1984)}]{Fischbach}%
  \BibitemOpen
  \bibfield  {author} {\bibinfo {author} {\bibfnamefont {E.}~\bibnamefont
  {Fischbach}}\ and\ \bibinfo {author} {\bibfnamefont {N.}~\bibnamefont
  {Nakagawa}},\ }\href {\doibase 10.1103/PhysRevD.30.2356} {\bibfield
  {journal} {\bibinfo  {journal} {Phys. Rev. D}\ }\textbf {\bibinfo {volume}
  {30}},\ \bibinfo {pages} {2356} (\bibinfo {year} {1984})}\BibitemShut
  {NoStop}%
\bibitem [{\citenamefont {Svozil}(1985)}]{Svozil}%
  \BibitemOpen
  \bibfield  {author} {\bibinfo {author} {\bibfnamefont {K.}~\bibnamefont
  {Svozil}},\ }\href {\doibase 10.1103/PhysRevLett.54.742} {\bibfield
  {journal} {\bibinfo  {journal} {Phys. Rev. Lett.}\ }\textbf {\bibinfo
  {volume} {54}},\ \bibinfo {pages} {742} (\bibinfo {year} {1985})}\BibitemShut
  {NoStop}%
\bibitem [{\citenamefont {Bordag}(1986)}]{Bordag}%
  \BibitemOpen
  \bibfield  {author} {\bibinfo {author} {\bibfnamefont {M.}~\bibnamefont
  {Bordag}},\ }\href {\doibase 10.1016/0370-2693(86)91009-9} {\bibfield
  {journal} {\bibinfo  {journal} {Phys. Lett. B}\ }\textbf {\bibinfo {volume}
  {171}},\ \bibinfo {pages} {113 } (\bibinfo {year} {1986})}\BibitemShut
  {NoStop}%
\bibitem [{\citenamefont {Boulware}\ \emph {et~al.}(1985)\citenamefont
  {Boulware}, \citenamefont {Brown},\ and\ \citenamefont
  {Lee}}]{BoulwareBrown}%
  \BibitemOpen
  \bibfield  {author} {\bibinfo {author} {\bibfnamefont {D.~G.}\ \bibnamefont
  {Boulware}}, \bibinfo {author} {\bibfnamefont {L.~S.}\ \bibnamefont {Brown}},
  \ and\ \bibinfo {author} {\bibfnamefont {T.}~\bibnamefont {Lee}},\ }\href
  {\doibase 10.1103/PhysRevD.32.729} {\bibfield  {journal} {\bibinfo  {journal}
  {Phys. Rev. D}\ }\textbf {\bibinfo {volume} {32}},\ \bibinfo {pages} {729}
  (\bibinfo {year} {1985})}\BibitemShut {NoStop}%
\bibitem [{\citenamefont {Kreuzer}\ and\ \citenamefont
  {Svozil}(1986)}]{KreuzerSvozil}%
  \BibitemOpen
  \bibfield  {author} {\bibinfo {author} {\bibfnamefont {M.}~\bibnamefont
  {Kreuzer}}\ and\ \bibinfo {author} {\bibfnamefont {K.}~\bibnamefont
  {Svozil}},\ }\href {\doibase 10.1103/PhysRevD.34.1429} {\bibfield  {journal}
  {\bibinfo  {journal} {Phys. Rev. D}\ }\textbf {\bibinfo {volume} {34}},\
  \bibinfo {pages} {1429} (\bibinfo {year} {1986})}\BibitemShut {NoStop}%
\bibitem [{\citenamefont {Kreuzer}(1988)}]{Kreuzer}%
  \BibitemOpen
  \bibfield  {author} {\bibinfo {author} {\bibfnamefont {M.}~\bibnamefont
  {Kreuzer}},\ }\href {\doibase 10.1088/0305-4470/21/15/017} {\bibfield
  {journal} {\bibinfo  {journal} {J. Phys. A: Math. Gen.}\ }\textbf {\bibinfo
  {volume} {21}},\ \bibinfo {pages} {3285} (\bibinfo {year}
  {1988})}\BibitemShut {NoStop}%
\bibitem [{\citenamefont {Barton}\ and\ \citenamefont
  {Fawcett}(1988)}]{BartonFawcett}%
  \BibitemOpen
  \bibfield  {author} {\bibinfo {author} {\bibfnamefont {G.}~\bibnamefont
  {Barton}}\ and\ \bibinfo {author} {\bibfnamefont {N.}~\bibnamefont
  {Fawcett}},\ }\href@noop {} {\bibfield  {journal} {\bibinfo  {journal}
  {Physics Reports}\ }\textbf {\bibinfo {volume} {170}},\ \bibinfo {pages} {1}
  (\bibinfo {year} {1988})}\BibitemShut {NoStop}%
\bibitem [{\citenamefont {Bennett}\ and\ \citenamefont
  {Eberlein}(2011)}]{MagMomPaper}%
  \BibitemOpen
  \bibfield  {author} {\bibinfo {author} {\bibfnamefont {R.}~\bibnamefont
  {Bennett}}\ and\ \bibinfo {author} {\bibfnamefont {C.}~\bibnamefont
  {Eberlein}},\ }\href@noop {} {\bibfield  {journal} {\bibinfo  {journal}
  {arXiv:1112.3224v1 [quant-ph]}\ } (\bibinfo {year} {2011})}\BibitemShut
  {NoStop}%
\bibitem [{\citenamefont {Eberlein}\ and\ \citenamefont
  {Robaschik}(2004)}]{EberleinRobaschik}%
  \BibitemOpen
  \bibfield  {author} {\bibinfo {author} {\bibfnamefont {C.}~\bibnamefont
  {Eberlein}}\ and\ \bibinfo {author} {\bibfnamefont {D.}~\bibnamefont
  {Robaschik}},\ }\href {\doibase 10.1103/PhysRevLett.92.233602} {\bibfield
  {journal} {\bibinfo  {journal} {Phys. Rev. Lett.}\ }\textbf {\bibinfo
  {volume} {92}},\ \bibinfo {pages} {233602} (\bibinfo {year}
  {2004})}\BibitemShut {NoStop}%
\bibitem [{\citenamefont {Eberlein}\ and\ \citenamefont
  {Robaschik}(2006)}]{PRDEberleinRobaschik}%
  \BibitemOpen
  \bibfield  {author} {\bibinfo {author} {\bibfnamefont {C.}~\bibnamefont
  {Eberlein}}\ and\ \bibinfo {author} {\bibfnamefont {D.}~\bibnamefont
  {Robaschik}},\ }\href {\doibase 10.1103/PhysRevD.73.025009} {\bibfield
  {journal} {\bibinfo  {journal} {Phys. Rev. D}\ }\textbf {\bibinfo {volume}
  {73}},\ \bibinfo {pages} {025009} (\bibinfo {year} {2006})}\BibitemShut
  {NoStop}%
\bibitem [{\citenamefont {Peskin}\ and\ \citenamefont
  {Schroeder}(1995)}]{Peskin}%
  \BibitemOpen
  \bibfield  {author} {\bibinfo {author} {\bibfnamefont {M.~E.}\ \bibnamefont
  {Peskin}}\ and\ \bibinfo {author} {\bibfnamefont {D.~V.}\ \bibnamefont
  {Schroeder}},\ }\href@noop {} {\emph {\bibinfo {title} {An Introduction to
  Quantum Field Theory}}}\ (\bibinfo  {publisher} {Westview Press},\ \bibinfo
  {address} {Boulder},\ \bibinfo {year} {1995})\BibitemShut {NoStop}%
\bibitem [{Note1()}]{Note1}%
  \BibitemOpen
  \bibinfo {note} {We work in natural units $c = 1 = \hbar $, $\epsilon _0 =
  1$.}\BibitemShut {Stop}%
\bibitem [{\citenamefont {Barton}(1977)}]{BartonPlasma1977}%
  \BibitemOpen
  \bibfield  {author} {\bibinfo {author} {\bibfnamefont {G.}~\bibnamefont
  {Barton}},\ }\href@noop {} {\bibfield  {journal} {\bibinfo  {journal} {J.
  Phys. A: Math. Gen.}\ }\textbf {\bibinfo {volume} {10}},\ \bibinfo {pages}
  {601} (\bibinfo {year} {1977})}\BibitemShut {NoStop}%
\bibitem [{\citenamefont {Babiker}(1976)}]{BabikerPlasma}%
  \BibitemOpen
  \bibfield  {author} {\bibinfo {author} {\bibfnamefont {M.}~\bibnamefont
  {Babiker}},\ }\href {\doibase 10.1103/PhysRevB.13.3056} {\bibfield  {journal}
  {\bibinfo  {journal} {Phys. Rev. B}\ }\textbf {\bibinfo {volume} {13}},\
  \bibinfo {pages} {3056} (\bibinfo {year} {1976})}\BibitemShut {NoStop}%
\bibitem [{\citenamefont {Glauber}\ and\ \citenamefont
  {Lewenstein}(1991)}]{GlauberLewenstein}%
  \BibitemOpen
  \bibfield  {author} {\bibinfo {author} {\bibfnamefont {R.~J.}\ \bibnamefont
  {Glauber}}\ and\ \bibinfo {author} {\bibfnamefont {M.}~\bibnamefont
  {Lewenstein}},\ }\href {\doibase 10.1103/PhysRevA.43.467} {\bibfield
  {journal} {\bibinfo  {journal} {Phys. Rev. A}\ }\textbf {\bibinfo {volume}
  {43}},\ \bibinfo {pages} {467} (\bibinfo {year} {1991})}\BibitemShut
  {NoStop}%
\bibitem [{\citenamefont {Scheel}\ and\ \citenamefont
  {Buhmann}(2008)}]{JenaReview}%
  \BibitemOpen
  \bibfield  {author} {\bibinfo {author} {\bibfnamefont {S.}~\bibnamefont
  {Scheel}}\ and\ \bibinfo {author} {\bibfnamefont {S.~Y.}\ \bibnamefont
  {Buhmann}},\ }\href@noop {} {\bibfield  {journal} {\bibinfo  {journal} {Acta
  Physica Slovaca}\ }\textbf {\bibinfo {volume} {58}},\ \bibinfo {pages} {675}
  (\bibinfo {year} {2008})}\BibitemShut {NoStop}%
\bibitem [{\citenamefont {Barton}(1997)}]{BartonPlasma}%
  \BibitemOpen
  \bibfield  {author} {\bibinfo {author} {\bibfnamefont {G.}~\bibnamefont
  {Barton}},\ }\href {\doibase 10.1098/rspa.1997.0132} {\bibfield  {journal}
  {\bibinfo  {journal} {Proc. R. Soc. Lond. A}\ }\textbf {\bibinfo {volume}
  {453}},\ \bibinfo {pages} {2461} (\bibinfo {year} {1997})}\BibitemShut
  {NoStop}%
\bibitem [{\citenamefont {Wu}\ and\ \citenamefont {Eberlein}(1999)}]{Wu}%
  \BibitemOpen
  \bibfield  {author} {\bibinfo {author} {\bibfnamefont {S.~T.}\ \bibnamefont
  {Wu}}\ and\ \bibinfo {author} {\bibfnamefont {C.}~\bibnamefont {Eberlein}},\
  }\href@noop {} {\bibfield  {journal} {\bibinfo  {journal} {Proc. R. Soc.
  Lond. A}\ }\textbf {\bibinfo {volume} {455}},\ \bibinfo {pages} {2487}
  (\bibinfo {year} {1999})}\BibitemShut {NoStop}%
\bibitem [{\citenamefont {Casimir}\ and\ \citenamefont {Polder}(1948)}]{CP}%
  \BibitemOpen
  \bibfield  {author} {\bibinfo {author} {\bibfnamefont {H.~B.~G.}\
  \bibnamefont {Casimir}}\ and\ \bibinfo {author} {\bibfnamefont
  {D.}~\bibnamefont {Polder}},\ }\href {\doibase 10.1103/PhysRev.73.360}
  {\bibfield  {journal} {\bibinfo  {journal} {Phys. Rev.}\ }\textbf {\bibinfo
  {volume} {73}},\ \bibinfo {pages} {360} (\bibinfo {year} {1948})}\BibitemShut
  {NoStop}%
\bibitem [{\citenamefont {Babiker}\ and\ \citenamefont
  {Barton}(1976)}]{BabikerBarton}%
  \BibitemOpen
  \bibfield  {author} {\bibinfo {author} {\bibfnamefont {M.}~\bibnamefont
  {Babiker}}\ and\ \bibinfo {author} {\bibfnamefont {G.}~\bibnamefont
  {Barton}},\ }\href@noop {} {\bibfield  {journal} {\bibinfo  {journal} {J.
  Phys. A: Math. Gen.}\ }\textbf {\bibinfo {volume} {9}},\ \bibinfo {pages}
  {129} (\bibinfo {year} {1976})}\BibitemShut {NoStop}%
\bibitem [{\citenamefont {Bordag}(2011)}]{BordagDrude}%
  \BibitemOpen
  \bibfield  {author} {\bibinfo {author} {\bibfnamefont {M.}~\bibnamefont
  {Bordag}},\ }\href@noop {} {\bibfield  {journal} {\bibinfo  {journal} {The
  European Physical Journal C - Particles and Fields}\ }\textbf {\bibinfo
  {volume} {71}},\ \bibinfo {pages} {1} (\bibinfo {year} {2011})}\BibitemShut
  {NoStop}%
\bibitem [{\citenamefont {Decca}\ \emph {et~al.}(2007)\citenamefont {Decca},
  \citenamefont {L\'opez}, \citenamefont {Fischbach}, \citenamefont
  {Klimchitskaya}, \citenamefont {Krause},\ and\ \citenamefont
  {Mostepanenko}}]{DeccaCasimirPressure}%
  \BibitemOpen
  \bibfield  {author} {\bibinfo {author} {\bibfnamefont {R.~S.}\ \bibnamefont
  {Decca}}, \bibinfo {author} {\bibfnamefont {D.}~\bibnamefont {L\'opez}},
  \bibinfo {author} {\bibfnamefont {E.}~\bibnamefont {Fischbach}}, \bibinfo
  {author} {\bibfnamefont {G.~L.}\ \bibnamefont {Klimchitskaya}}, \bibinfo
  {author} {\bibfnamefont {D.~E.}\ \bibnamefont {Krause}}, \ and\ \bibinfo
  {author} {\bibfnamefont {V.~M.}\ \bibnamefont {Mostepanenko}},\ }\href
  {\doibase 10.1103/PhysRevD.75.077101} {\bibfield  {journal} {\bibinfo
  {journal} {Phys. Rev. D}\ }\textbf {\bibinfo {volume} {75}},\ \bibinfo
  {pages} {077101} (\bibinfo {year} {2007})}\BibitemShut {NoStop}%
\end{thebibliography}
\end{document}